\documentclass[10pt,conference]{IEEEtran}
\IEEEoverridecommandlockouts
\usepackage{cite}
\usepackage{amsmath,amssymb,amsfonts}
\usepackage{algorithmic}
\usepackage{booktabs}
\usepackage{graphicx}
\usepackage{soul}
\usepackage{textcomp}
\usepackage{listings}
\usepackage{xcolor}
\usepackage{lipsum}
\usepackage{textcomp}
\usepackage{multirow}
\usepackage{tabularx}
\newcommand{\ket}[1]{\left| {#1} \right\rangle}

\newcommand{\tket}{$\mathrm{t\ket{ket}}$}

\definecolor{codegreen}{rgb}{0,0.6,0}
\definecolor{codegray}{rgb}{0.5,0.5,0.5}
\definecolor{codepurple}{rgb}{0.58,0,0.82}
\definecolor{backcolour}{rgb}{0.97,0.97,0.95}

\lstdefinestyle{mystyle}{
    backgroundcolor=\color{backcolour},   
    commentstyle=\color{codegreen},
    keywordstyle=\color{magenta},
    numberstyle=\tiny\color{codegray},
    stringstyle=\color{codepurple},
    basicstyle=\ttfamily\scriptsize
    ,
    breakatwhitespace=false,         
    breaklines=true,                 
    captionpos=b,                    
    keepspaces=true,                 
    numbers=left,                    
    numbersep=5pt,                  
    showspaces=false,                
    showstringspaces=false,
    showtabs=false,                  
    tabsize=2
}

\def\BibTeX{{\rm B\kern-.05em{\sc i\kern-.025em b}\kern-.08em
    T\kern-.1667em\lower.7ex\hbox{E}\kern-.125emX}}
\begin{document}

\title{Qsyn: A Developer-Friendly Quantum Circuit Synthesis Framework for NISQ Era and Beyond\\

\thanks{The authors would like to thank Prof. Hao-Chung Cheng for his invaluable feedback from the users' perspectives during the development of Qsyn. This work is supported by the National Science and Technology Council, Taiwan. Project No.: NSTC 112-2119-M-002-017.}
}

\author{\IEEEauthorblockN{
Mu-Te Lau\IEEEauthorrefmark{1}\IEEEauthorrefmark{2}, 
Chin-Yi Cheng\IEEEauthorrefmark{1}\IEEEauthorrefmark{2}, 
Cheng-Hua Lu\IEEEauthorrefmark{1}\IEEEauthorrefmark{3}, 
Chia-Hsu Chuang\IEEEauthorrefmark{1}\IEEEauthorrefmark{2}, 
\\
Yi-Hsiang Kuo\IEEEauthorrefmark{2}, 
Hsiang-Chun Yang\IEEEauthorrefmark{2}, 
Chien-Tung Kuo\IEEEauthorrefmark{2}, 
Hsin-Yu Chen\IEEEauthorrefmark{2}, 
Chen-Ying Tung\IEEEauthorrefmark{2}, 
Cheng-En Tsai\IEEEauthorrefmark{2}, 
\\
Guan-Hao Chen\IEEEauthorrefmark{2}, 
Leng-Kai Lin\IEEEauthorrefmark{2}, 
Ching-Huan Wang\IEEEauthorrefmark{2}, 
Tzu-Hsu Wang\IEEEauthorrefmark{2}, 
Chung-Yang Ric Huang\IEEEauthorrefmark{2}}
\\

  \IEEEauthorblockA{\IEEEauthorrefmark{2}National Taiwan University, R.O.C., \IEEEauthorrefmark{3}Stanford University, U.S.}
  \IEEEauthorblockA{josh.lau2011@gmail.com, chin-yi.cheng@utexas.edu, cyhuang@ntu.edu.tw}
}

\maketitle
\pagestyle{plain}

\begin{abstract}
In this paper, we introduce a new quantum circuit synthesis (QCS) framework, Qsyn, for developers to research, develop, test, experiment, and then contribute their QCS algorithms and tools to the framework. Our framework is more developer-friendly than other modern QCS frameworks in three aspects: (1) We design a rich command-line interface so that developers can easily design various testing scenarios and flexibly conduct experiments on their algorithms. (2) We offer detailed access to many data representations on different abstract levels of quantum circuits so that developers can optimize their algorithms to the extreme. (3) We define a rigid developing flow and environment so that developers can ensure their development qualities with the best modern software engineering practices. We illustrate the friendliness of our framework with a showcase of developing a T-Count Optimization algorithm and demonstrate our performance superiority with fair comparisons to other modern QCS frameworks.
\end{abstract}


\section{Introduction} \label{section:introduction}
Quantum circuit synthesis (QCS) is indispensable for realizing quantum advantage. A QCS flow transforms high-level descriptions of quantum algorithms into low-level instructions directly executable on quantum hardware. A well-designed QCS process is crucial to unlocking a quantum system's full potential. It facilitates more efficient computation, greater error tolerance, and, ultimately, a stronger quantum advantage. 

To realize this envision, past research has proposed several QCS algorithms, with many of them also publishing corresponding tools \cite{soeken_epfl_2018, campbell_efficient_2016, de_beaudrap_fast_2020, wille_mqt_2023, ross_optimal_2016, meuli_reversible_2019, meuli_ros_2020, soeken_revkit_2012, amy_polynomial-time_2014, heyfron_efficient_2018, bravyi_clifford_2021, kissinger_reducing_2020}. However, most of these tools focused on particular features and did not provide friendly user experiences. What is worse, the different input/output formats and varying programming languages hamper their interoperability with each other.

IBM has published Qiskit~\cite{aleksandrowicz_qiskit_2019}, one of the first frameworks that provides an end-to-end QCS flow. Its transpiler converts the input quantum circuit to make it executable on IBM's own quantum devices~\cite{Ibmq2021}. Other synthesis frameworks, such as Microsoft's Q\#~\cite{microsoft_q_2020}, Rigetti's PyQuil~\cite{smith_practical_2017}, 
Google's Cirq~\cite{cirq_developers_cirq_2023}, and numerous others \cite{heurtel_perceval_2023, killoran_strawberry_2019, bromley_applications_2020}, also provide similar flows, often tailored to their respective quantum devices. On the other hand, projects such as Quantinuum's \tket~\cite{sivarajah_tket_2021, cowtan_phase_2020}, SoftwareQ Inc's \texttt{staq} \cite{amy_staq_2020}, PyZX \cite{kissinger_pyzx_2020}, and many more \cite{green_quipper_2013, javadiabhari_scaffcc_2014, bichsel_silq_2020} focus on offering general QCS functionalities that do not tie to any particular devices. They derive a common hardware-agnostic representation before converting to backends.

While these frameworks provide complete features for QCS, their primary target users are quantum algorithm/circuit designers, who focus on utilizing these existing tools to implement quantum algorithms and/or synthesize them into quantum circuits. However, to push the advancement of the QCS algorithms in the future, we need a framework that offers a friendly environment for more QCS algorithm/tool developers (called ``developers'' in the following) to easily contribute their advanced ideas and conduct thorough experiments. In contrast to the existing QCS frameworks, such a ``developer-friendly'' framework should provide the following features:

\begin{enumerate}
    \item \emph{A unified and convenient platform to conduct experiments.}
    Currently, QCS algorithms are typically implemented from scratch, using various programming languages and underlying data structures. This severely complicates the assessment of the algorithms' runtime and memory efficiency and drags development since new developers routinely spend time reinventing the wheel.
    \item \emph{An interface to access low-level data directly.}
    While existing frameworks provide easy-to-use QCS functionalities for end users, developers of QCS algorithms would greatly benefit from the ability to inspect and analyze the algorithms' behavior in runtime.
\end{enumerate}

We have open-sourced Qsyn, a developer-friendly QCS framework, to aid further development in this field. Our main contributions are:
\begin{enumerate}
    \item Providing a unified and user-friendly developing environment so researchers and developers can efficiently prototype, implement, and thoroughly evaluate their QCS algorithms with standardized tools, language, and data structures.
    \item Assisting the developers with a robust and intuitive interface that can access low-level data directly to provide developers with unique insights into the behavior of their algorithms during runtime.
    \item Enforcing robust quality-assurance practices, such as regression tests, continuous integration-and-continuous delivery (CI/CD) flows, linting, etc. These methodologies ensure that we provide reliable and efficient functionalities and that new features adhere to the same quality we strive for. 
\end{enumerate}

With Qsyn, developers can easily accelerate the implementation and evaluation of new QCS algorithms by leveraging the provided data structures and development environments. Our series of efforts are inspired by ABC~\cite{hutchison_abc_2010} and Yosys~\cite{wolf_yosys-free_2013}, two logic synthesis and verification frameworks widely adopted in the field of electronic design automation (EDA). 

The rest of this paper is organized as follows. Section~\ref{section:background} introduces the background knowledge on the QCS problem. Section~\ref{section:functionalities} discusses the main functionalities of Qsyn, and Section~\ref{section:highlight} highlights our advantage over other similar frameworks. We demonstrate in Section~\ref{section:casestudy} a typical workflow of developing QCS algorithms in Qsyn, and detail in Section~\ref{section:experiments} the experimental evaluation of our synthesis framework. Section~\ref{section:conclusion} wraps up our work and gives future directions.
\section{Background} \label{section:background}
Quantum circuit synthesis (QCS) is a multiple-step process that transforms high-level quantum algorithms into optimized and executable circuits for quantum computing devices. In the burgeoning field of quantum computing, the challenge of this task is accentuated by the ever-evolving quantum device architectures. In this context, we first explore QCS for noisy intermediate-scale quantum (NISQ) devices before describing the challenges of moving beyond NISQ devices. Then, we discuss the strategic structuring of QCS frameworks to foster development in this rapidly evolving domain.

\subsection{Quantum circuit synthesis}
\begin{figure}
    \centering
    \includegraphics[width=0.8\linewidth]{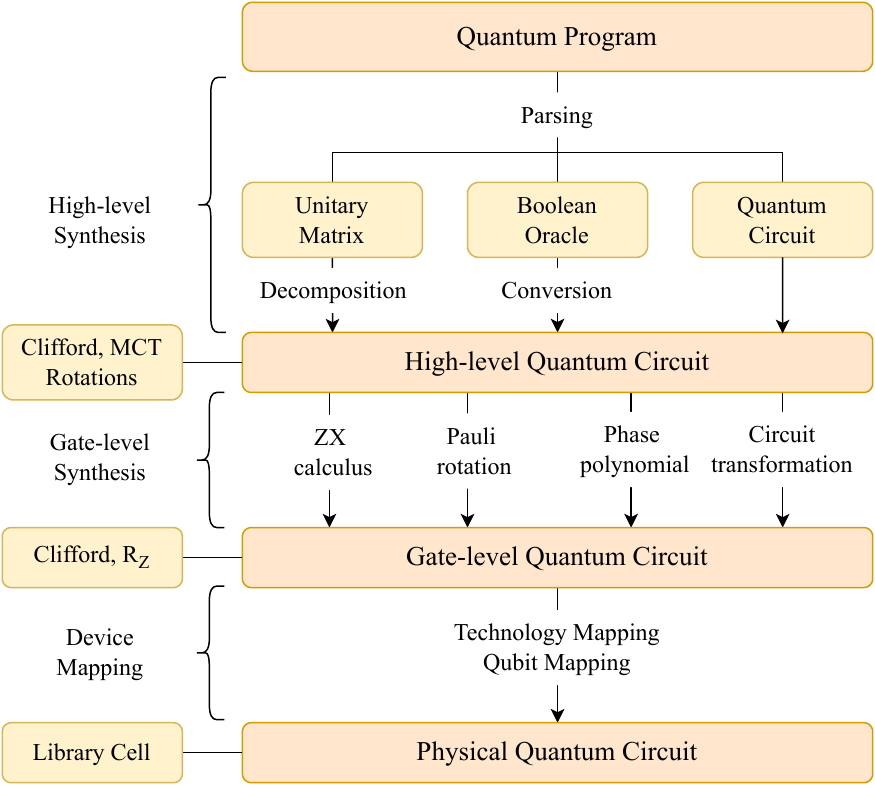}
    \caption{A typical quantum circuit synthesis flow.}
    \vspace{-10px}
    \label{fig:compilation-flow}
\end{figure}

Fig.~\ref{fig:compilation-flow} illustrates a typical QCS flow. The process starts by parsing a high-level quantum circuit from an abstract language, then synthesizing it into basic gate types, and finally adapting it to the target quantum device's gate set and connectivity.

\subsubsection{High-level synthesis}
Quantum algorithms are usually represented as quantum circuits with large and complex components, such as Boolean oracles and unitary matrices. To execute these components on quantum devices and simplify optimization processes at later stages, they are first synthesized into simpler quantum gates, such as multiple-controlled Toffoli (MCT) gates, Clifford gates, and single-qubit rotation gates. 

While the synthesis methods vary based on the type of components, they all typically produce a large number of MCT gates. Therefore, the goal at this stage is to reduce the number of MCT gates, especially those with many controls, as their low-level implementation includes using numerous small-angle rotation gates and/or ancillae, which significantly increases the circuit size.

\subsubsection{Gate-level logical circuit synthesis}
The gate-level synthesis stage produces optimized quantum circuits containing only fundamental gates. A common target is the Clifford+$R_Z$ gate set generated by $\{ CX, H, R_Z(\theta) \}$. Common synthesis approaches include ZX-calculus-based \cite{ kissinger_reducing_2020, staudacher_reducing_2023}, Pauli rotation-based \cite{zhang_optimizing_2019, vandaele_optimal_2024}, phase-polynomial-based \cite{heyfron_efficient_2018, de_beaudrap_fast_2020, amy_polynomial-time_2014}, and circuit transformation-based methodologies \cite{bravyi_clifford_2021, xu_quartz_2022}, etc. The first two methods address Clifford+$R_Z$ circuits, while the phase polynomial-based can only address Clifford+$T$ circuits but can achieve even better optimization results in exchange. The commonality of these three approaches is that they all transform the quantum circuit into a certain data representation. In contrast, the circuit transformation-based methods directly work on the netlist of the quantum circuit. The power and limitations of the circuit transformation-based approach vary significantly between algorithms.

Traditionally, metrics for evaluating circuit performance include the number of $T$ gates or non-Clifford rotation gates, i.e., $T$-count and $R_Z$-count, for they are costly to implement on proposed fault-tolerant quantum devices~\cite{campbell_efficient_2016, litinski_magic_2019}. However, recent research has emphasized other metrics, such as circuit depths, $H$ counts~\cite{vandaele_optimal_2024}, and two-qubit gate counts~\cite{staudacher_reducing_2023}, which help achieve smaller circuit sizes and lower execution times.

\subsubsection{Device mapping}
The final stage of QCS involves adapting the optimized circuit to a specific quantum device, ensuring its compatibility with the device's native gate set and qubit connectivity. This stage primarily contains two tasks: technology mapping and qubit mapping.

Technology mapping involves converting a circuit to the target device's native gate set, a set of gates natively supported by a specific device. 
For example, IBM's early quantum devices adopt $\{ CX, R_Z(\theta), S_X, X, I \}$, while more recent devices replace the $CX$ gate with the echoed cross-resonance gate ($ECR$) for better operational accuracy~\cite{Ibmq2021, sundaresan_reducing_2020}. 

On the other hand, qubit mapping aims to conform a quantum circuit to the quantum devices' connectivity because two-qubit gates are applicable only to adjacent pairs of physical qubits. 
On NISQ devices, the qubit mapping problem involves assigning the logical qubits to physical ones and inserting necessary SWAP gates to enable two-qubit operations. 

The primary objective of the device mapping stage is to optimize circuit fidelity. For smaller quantum circuits, fidelity measurement serves as a direct indicator of reliability on the target device. However, for larger circuits, where direct fidelity measurement is impractical, it is common to adopt other metrics, such as counting the number of inserted SWAP gates or evaluating the depth of the mapped quantum circuits.

\vspace{-5px}
\subsection{Synthesis for devices beyond NISQ era}
The need for advanced synthesis strategies becomes critical as quantum computing technology evolves beyond near-intermediate-scale quantum (NISQ) devices, including the consideration of fault-tolerant mechanisms, distributive computations, and low-level control of devices.

\subsubsection{Fault-tolerant architectures}
Research has proposed integrating a quantum error correction (QEC) layer for fault-tolerant, large-scale quantum devices. This incorporation fundamentally modifies the QCS problem due to distinct gate implementations in fault-tolerant settings. 

For instance, although arbitrary $Z$-axis rotations ($R_Z(\theta)$) are relatively simple on NISQ devices, their implementation in fault-tolerant systems usually involves complex techniques, such as state distillation \cite{campbell_efficient_2016, litinski_magic_2019}, and thereby the resource demand significantly increases as the rotation angles get finer. As a result, circuits must be adapted to gate sets like Clifford+$T$, i.e., $\{ CX, H, T \}$~\cite{dawson_solovay-kitaev_2006, 
ross_optimal_2016}. Additionally, QEC utilizes unique protocols, such as lattice surgery or double-defect braiding, to implement Clifford gates while maintaining the robustness of the QEC~\cite{horsman_surface_2012, fowler_surface_2012}. These methods impose specific connectivity and locality requirements that influence the QCS algorithm design.

\subsubsection{Distributive quantum devices}
Due to physical and engineering constraints, centralized quantum processors face significant challenges concerning scalability and error rates. Leveraging on entanglements, distributive quantum devices present a feasible alternative by distributing computational tasks across multiple nodes, potentially enhancing the fault tolerance and reducing the quantum error rate~\cite{tang_cutqc_2021, zhang_mech_2023, mao_qubit_2023, davis_towards_2023, escofet_revisiting_2024}.
Some budding compilation frameworks also target the distributed quantum compilation paradigm~\cite{wu_autocomm_2022, ferrari_modular_2023}.

\subsubsection{Low-level control}
In addition to scaling up the device sizes, research has proposed device-specific strategies to maximize the capabilities of NISQ devices by considering their low-level characteristics. For example, direct pulse control of superconducting devices~\cite{shi_optimized_2019, gokhale_optimized_2020} can significantly shorten execution time, a critical factor for operations in limited coherence. Moreover, Echoed Cross-Resonance (ECR) mitigates unwanted crosstalking and dephasing, improving the gate fidelity in superconducting NISQ devices~\cite{sundaresan_reducing_2020}. In addition to exploiting existing technologies, these approaches also make the existing device more robust, aiding in the development of distributive quantum devices.

\subsection{QCS framework for NISQ era and beyond}

Looking from the perspectives of scalability and applicability, going beyond the NISQ era in quantum computing is analogous to the advent of the VLSI era in classical electronic computing. Therefore, to construct a QCS framework for the NISQ era and beyond, it is plausible for us to learn from the history of the successful development of the VLSI industry.

To the best of our knowledge, Electronics Design Automation (EDA) was the key enabler for the rapid growth of VLSI technologies in the past decades. However, in addition to the EDA tools from various EDA vendors, open-source frameworks also played an essential role in these advancements. 

Particularly in the logic synthesis of the classical circuits, MIS from UC Berkeley \cite{brayton_mis_1987} and follow-up frameworks such as SIS \cite{sentovich_sis_1992}, VIS \cite{goos_vis_1996} and ABC \cite{hutchison_abc_2010} have inspired numerous researchers to participate and contribute in this area, even nurturing many advanced commercial tools and synthesis startups. It is fair to acknowledge that these open-source frameworks were the mothers of modern logic synthesis technologies.

In short, as we are moving beyond the NISQ era in the foreseeable future, what we need for QCS will not just be proprietary tools or techniques but a developer-friendly framework so that more ideas and experiments can be realized on top of it with ease. Hence, in this subsection, we will review the basic, but crucial aspects of constructing a developer-friendly framework. It will help establish the cornerstones of the QCS framework for the NISQ era and beyond.



\subsubsection{Programmable and extensible user interface for algorithm designs and experiments}
The first essential feature of a developer-friendly framework is an easily programmable and extensible user interface so that developers can design varied scenarios for the algorithms and experiments. One common approach among QCS frameworks is providing a library, e.g., IBM's Qiskit~\cite{aleksandrowicz_qiskit_2019}, where developers can leverage synthesis APIs to program to their needs. They can also extend Qiskit's capability by contributing new algorithms.
Another common approach is devising a compiler for quantum programming languages. such as ScaffCC~\cite{javadiabhari_scaffcc_2014} for the Scaffold language~\cite{javadiabhari_scaffold_2012}. Developers can program the compiler's behavior by supplying optimization passes as compiler flags or extend the compiler's ability by supplying new algorithms as flags.

The library-based approach excels at programmability and extensibility by adopting popular programming languages that developers can swiftly maneuver. However, whether developers can fully grasp the framework heavily depends on the quality of the documentation. On the other hand, programming and extending a compiler is not as easy, though developers may learn the available synthesis strategies quicker through the more centralized compiler options.

\subsubsection{Diverse data representations with secure low-level access and operation}
Since QCS is a multi-stage process that transforms high-level quantum algorithms into optimized and executable circuits for quantum computing devices, there will be diverse data representations for different stages, algorithms, and even hardware technologies. For example, in high-level synthesis, we need multiple-controlled Toffoli (MCT) gates to represent Boolean oracles and unitary matrices to describe the linear transformations of specific algorithms. For gate-level optimization, the data structure of the quantum gates should be able to handle various basic gate types for different quantum devices and can be converted to different logical structures, such as ZX-diagrams, Pauli rotations, phase polynomials, etc., for different synthesis algorithms.

More importantly, a developer-friendly QCS framework should offer secure low-level data access and operation during the runtime of the algorithms. Such a framework can boost troubleshooting and fine-tuning efficiencies in development. It also enables developers to probe, diagnose, and rectify discrepancies or bottlenecks instantly.



\subsubsection{Good mechanism to coalesce development efforts}
A well-established framework should be open to incorporating the efforts of external developers in order to build an ecosystem and participate in advancing the field. This requires a good mechanism to maintain the code quality and framework's stability so as to minimize the friction among developers and thus maximize the synergistic effects of the community.

A great example would be the MQT-QMAP project \cite{wille_mqt_2023}, which serves as a common platform for researchers of qubit mapping algorithms. It provides varied data structures and utilities so that numerous cutting-edge studies can be devised on MQT-QMAP. The QCS field, as a whole, would also benefit from having such a holistic development environment.

\section{Qsyn: a Developer-Friendly Quantum Circuit Synthesis Framework} \label{section:functionalities}
We present Qsyn\footnote{https://github.com/DVLab-NTU/qsyn}, an open-source QCS framework that provides a swift development experience of QCS algorithms. Qsyn aims to coalesce the efforts in the QCS field by providing unified data structures and a flexible command-line interface (CLI) so that developers can easily create new algorithms, implement them as commands, and benchmark them against existing approaches.
\vspace{-5px}
\subsection{Installation and usage} \label{subsection:functionalities:usage}
Qsyn is written with the up-to-date C++-20 standard to ensure developers can leverage modern programming practices, such as functional programming and range adaptors, to speed up development. Qsyn is easy to compile and adopts CMake for the developers to import other libraries and manage dependencies. 

To use Qsyn, the user just downloads the source code from the GitHub repository and compiles it by running \texttt{make} in the project root directory, which triggers our suggested CMake compilation flow. After successful compilation, the user can run the \texttt{./qsyn} binary and be greeted by a command-line interface
where they can type in commands. A good way to start navigating would be using \texttt{help} to list all available commands:
\begin{lstlisting}
qsyn> help
\end{lstlisting}
Additionally, appending the \texttt{-h} or \texttt{--help} flag behind any command will print out its usage and detailed explanation. For example, the user can learn how to use the \texttt{qcir read} command by invoking the following command call:
\begin{lstlisting}
qsyn> qcir read -h
Usage: qcir read [-h] [-r] <string filepath>

Description:
  read a quantum circuit and construct the 
  corresponding netlist

Positional Arguments:
  string  filepath    the filepath to the quantum
                      circuit file. Supported 
                      extension: .qasm, .qc 

Options:
  flag  -h, --help       show this help message                                                  
  flag  -r, --replace    if specified, replace the 
                         current circuit; otherwise 
                         store a new one 
\end{lstlisting}

Qsyn also supports reading commands from scripts. For example, the following script, which we have also provided in \texttt{examples/zxopt.qsyn} in the project folder, runs a ZX-calculus-based optimization flow:
\begin{lstlisting}[language=c]
//!ARGS INPUT
qcir read ${INPUT}
echo "--- pre-optimization  ---"
qcir print --stat
// to zx -> full reduction -> extract qcir
qc2zx; zx optimize --full; zx2qc
// post-resyn optimization
qcir optimize
echo "--- post-optimization ---"
qcir print --stat
\end{lstlisting}

In the above script, the banner \texttt{//!ARGS INPUT} mandates that the caller provide exactly one argument for the variable \texttt{INPUT}. Texts following \texttt{//} in the same line are comments to the scripts. To run the commands in the above file, supply the script's file path and arguments after \texttt{./qsyn}:
\begin{lstlisting}[language=bash]
$ ./qsyn examples/zxopt.qsyn \
         benchmark/SABRE/large/adr4_197.qasm
\end{lstlisting}
This runs the ZX-calculus-based synthesis on the circuit in \texttt{benchmark/SABRE/large/adr4\_197.qasm}. Developers can utilize scripts to automate repetitive tasks and use Qsyn along with other tools in the shell. We shall discuss the shell interoperability of Qsyn in more detail in Section~\ref{section:highlight}.
\vspace{-5px}
\subsection{Main functionalities} \label{subsection:functionalities:functionalities}
At the time of writing, Qsyn can perform end-to-end synthesis by invoking functionalities for each synthesis stage as commands. It also offers a series of data access and utility commands for developers to leverage. Below, we will briefly introduce each aspect of Qsyn's functionalities, complemented by some of the exemplar commands\footnote{Due to space limit, some of the commands are illustrated with aliases.}.

\subsubsection{High-level synthesis}
Qsyn can process various specifications for quantum circuits by supporting syntheses from Boolean oracles and unitary matrices:
\begin{center}
\footnotesize
\begin{tabularx}{\linewidth}{lXc}
    \toprule
        Command & Description & Ref  \\
    \midrule\midrule
        \texttt{qcir oracle} & ROS Boolean oracle synthesis flow & \cite{meuli_ros_2020, meuli_reversible_2019} \\
        \texttt{ts2qc} & Gray-code unitary matrix synthesis & \cite{krol_efficient_2022, nakahara_quantum_2008, shende_minimal_2004} \\
    \bottomrule
\end{tabularx}
\end{center}
\subsubsection{Gate-level synthesis}
Qsyn can adapt to different optimization targets by providing different routes to synthesize low-level quantum circuits. In the following table, \texttt{qzq} and \texttt{qtablq} are two gate-level synthesis routines, with the indented commands composing their algorithms.
\begin{center}
    \footnotesize
    \begin{tabularx}{\linewidth}{lXc}
    \toprule
        Command & Description & Ref  \\
    \midrule\midrule
        \texttt{qzq}                  & ZX-calculus-based synth routine       & \cite{kissinger_pyzx_2020} \\ 
        \texttt{\ \ qc2zx}            & convert q. circuit to ZX-diagram      & - \\
        \texttt{\ \ zx opt}           & fully reduce ZX diagram               & \cite{kissinger_reducing_2020} \\
        \texttt{\ \ zx2qc}            & convert ZX-diagram to q. circuit      & \cite{kissinger_reducing_2020} \\
        \texttt{\ \ qc opt}           & basic optimization passes             & \cite{kissinger_reducing_2020} \\
        \texttt{qtablq}               & tableau-based synth routine           & - \\
        \texttt{\ \ qc2tabl}          & convert quantum circuit to tableau    & - \\
        \texttt{\ \ tabl opt full}    & iteratively apply the following three & - \\
        \texttt{\ \ tabl opt tmerge}  & phase-merging optimization            & \cite{zhang_optimizing_2019} \\
        \texttt{\ \ tabl opt hopt}    & internal $H$-gate optimization        & \cite{vandaele_optimal_2024} \\
        \texttt{\ \ tabl opt ph todd} & TODD optimization                     & \cite{heyfron_efficient_2018} \\
        \texttt{\ \ tabl2qc}          & convert tableau to quantum circuit    & - \\
        \texttt{sk-decompose}         & Solovay-Kitaev decomposition & \cite{dawson_solovay-kitaev_2006}\\
    \bottomrule
    \end{tabularx}
\end{center}

\subsubsection{Device mapping}
Qsyn can target a wide variety of quantum devices by addressing their available gate sets and topological constraints.
\begin{center}
    \footnotesize
    \begin{tabularx}{\linewidth}{lXc}
    \toprule
        Command & Description & Ref  \\
    \midrule\midrule
        \texttt{device read}  & read info about a quantum device        & - \\
        \texttt{qc translate} & library-based technology mapping        & - \\
        \texttt{qc opt -t}    & technology-aware optimization passes    & - \\
        \texttt{duostra}      & Duostra qubit mapping for depth or \#SWAPs & \cite{cheng_robust_2024} \\
    \bottomrule
    \end{tabularx}
\end{center}

\subsubsection{Data access and utilities} 
Qsyn provides various data representations for quantum logic. We list the common commands to them in the following table. Here, \texttt{<dt>} stands for any data representation type, including quantum circuits, ZX-diagrams, Tableau, etc.:
\begin{center}
    \footnotesize
    \begin{tabularx}{\linewidth}{lX}
    \toprule
        Command & Description  \\
    \midrule\midrule
        \texttt{<dt> list}           & list all \texttt{<dt>}s       \\
        \texttt{<dt> checkout}       & switch focus between \texttt{<dt>}s       \\
        \texttt{<dt> print}          & print \texttt{<dt>} information     \\
        \texttt{<dt> new|delete}     & add a new/delete a \texttt{<dt>}         \\
        \texttt{<dt> read|write}     & read and write \texttt{<dt>}              \\
        \texttt{<dt> equiv}          & verify equivalence of two \texttt{<dt>}s   \\
        \texttt{<dt> draw}           & render visualization of \texttt{<dt>}     \\
        \texttt{convert <dt1> <dt2>} & convert from \texttt{<dt1>} to \texttt{<dt2>}   \\
    \bottomrule
    \end{tabularx}
\end{center}
Besides the above commands, Qsyn offers several utilities for benchmarking, debugging, and customizing:
\begin{center}
    \footnotesize
    \begin{tabularx}{\linewidth}{lX}
    \toprule
        Command & Description \\
    \midrule\midrule
        \texttt{alias}   & set or unset aliases           \\
        \texttt{help}    & display helps to commands      \\
        \texttt{history} & show or export command history \\
        \texttt{logger}  & control log levels             \\
        \texttt{set}     & set or unset variables         \\
        \texttt{usage}   & show time/memory usage         \\
    \bottomrule
    \end{tabularx}
\end{center}

\subsection{Software architecture}
\begin{figure}[t]
  \centering
  \includegraphics[width=\linewidth]{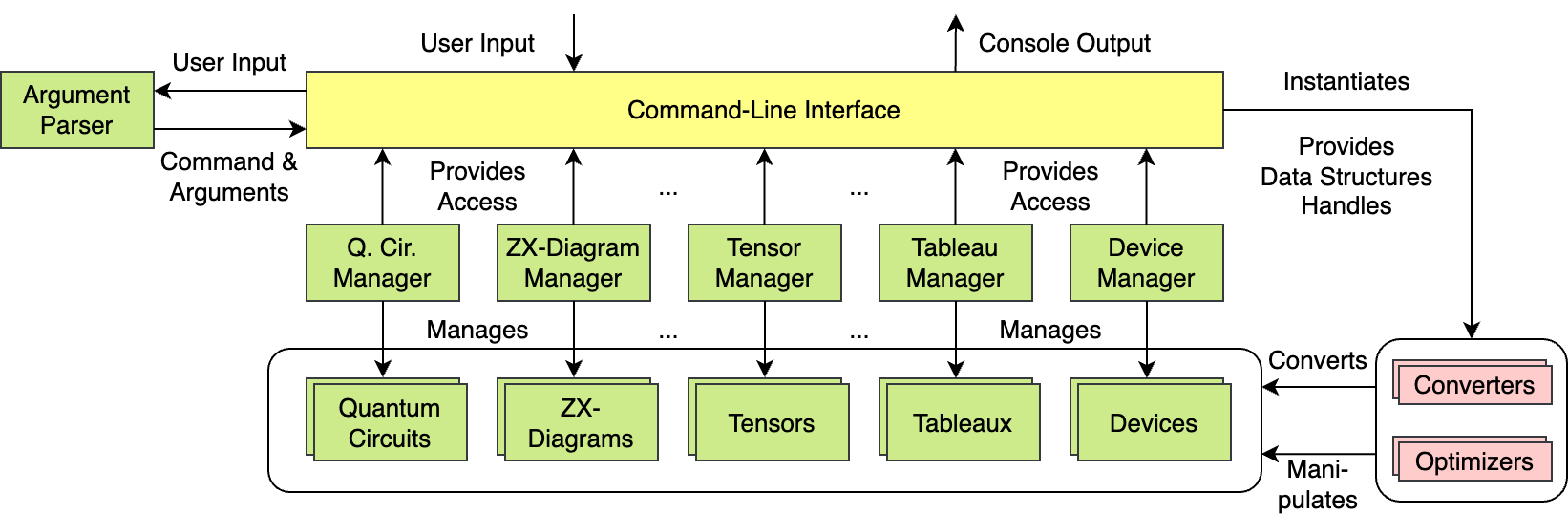}
  \vspace{-20px}
  \caption{Qsyn's software architecture.}
  \label{fig:qsyn-arch}
    \vspace{-10px}
\end{figure}

Fig.~\ref{fig:qsyn-arch} depicts the software architecture of Qsyn. The core interaction with Qsyn is facilitated through its command-line interface (CLI), which processes user input, handles command execution, and manages error reporting. Extensibility is achieved by allowing developers to integrate additional commands into the CLI. We provide further details in Section~\ref{section:casestudy}.

Qsyn is designed with extensibility in mind, leveraging a data-oriented approach focusing on robust data management and manipulation capabilities. It offers API for various quantum circuit representations like ZX diagrams and Tableaux. Higher-level operations like synthesis algorithms can be implemented with free functions or strategy classes. This design principle ensures that new strategies can be added without compromising existing data structures, as all modifications are funneled through well-defined public interfaces.

To enhance experiments and testing, Qsyn supports real-time storage of multiple quantum circuits and intermediate representations by the \texttt{<dt> checkout} command. Users can take snapshots at any time in the synthesis process and switch to an arbitrary version of stored data. This facilitates the design of the experiments by branching into varied testing scenarios on different snapshots. It also allows simultaneous access to multiple versions of circuits for advanced operations such as equivalence checking, circuit composition, etc.

This architecture is central to Qsyn's flexibility and extensibility. It segregates responsibilities and simplifies command implementations while enhancing its capability as a research tool in quantum circuit synthesis.

\section{Framework Highlights} \label{section:highlight}
This section presents the highlights of the Qsyn framework, including (1) Our command line interface (CLI) to assist developers in designing new algorithms, (2) Our data representations and utilities to offer convenient and advantageous implementations, and (3) Our efforts in project control and community engagement to ensure the sustainable development of the framework.

\subsection{Development of new algorithms}
Qsyn's mission is to assist developers in creating and assessing new synthesis algorithms. Below, we will sketch how developers may define the usage flows of their algorithms by leveraging the functionalities provided by our CLI.
\subsubsection{Powerful and easy-to-maintain argument parser} 
A usage flow in Qsyn is composed of commands. Qsyn streamlines the addition of a new command with its powerful argument parser. It relieves developers' burden in parsing complex commands and, at the same time, automatically generates up-to-date command manuals. The following code demonstrates the argument parser for the \texttt{qcir read} command in Qsyn:
\begin{lstlisting}[language=c++]
using namespace dvlab::argparse;
auto parser = ArgumentParser{}
  .description("read a quantum circuit")
parser.add_argument<std::string>("filepath")
  .help("path to circuit file");
parser.add_argument<bool>("-r", "--replace")
  .action(store_true)
  .help("replace the current circuit");
\end{lstlisting}

This above code illustrates an argument parser that defines a mandatory \texttt{filepath} argument and an optional \texttt{-r|--replace} flag. The API intentionally resembles the Python \texttt{argparse} package so developers can learn its usage effortlessly. A detailed demonstration of defining a command and its argument parser is covered in Section~\ref{section:casestudy}. By offering a standardized parser package, Qsyn ensures uniformity in command syntax, enhanced clarity in help manuals, and precise error messages, allowing developers to devote their full attention to research tasks.

\subsubsection{Useful CLI utilities}
Qsyn CLI supports up/down arrow keys to retrieve commands in history. Also, the \texttt{history} command allows printing or exporting of executed commands. This exporting functionality is particularly convenient, facilitating reapplication in relevant contexts. 

Personalization is a strong feature of Qsyn; with the \texttt{alias} and \texttt{set} commands, users can create aliases and variables to customize the CLI to their needs. For persistent use of these personalized elements, they can be added to the config file \texttt{\textasciitilde/.config/qsyn/qsynrc}. Notably, Qsyn optimizes user experience by enabling command, alias, or file path auto-completion and expansion of complete aliases or variables into full form with the Tab key.

\subsubsection{Verification and testing}
One crucial aspect of developing new synthesis algorithms is verifying their correctness. For small circuits ($\leq\!10$ qubits), Qsyn can directly convert the circuits into a tensor using \texttt{ts2qc} and checks for equivalence numerically with \texttt{tensor equiv}.

For larger circuits, symbolic verification can be applied by first appending the reversed copy of the circuit to its end. Then, we can invoke an optimization routine to try reducing the composed circuit to identity---or at least make it compact enough for easier numerical calculations. The effectiveness of this approach naturally depends on the robustness of the optimization routine.

After verifying the algorithm's validity, the next step is to assess its performance. Developers can use the \texttt{usage} command to gather essential data regarding the algorithm's runtime and memory usage. Furthermore, they can retrieve details about the final circuit statistics by employing the \texttt{qcir print --stat} command, which provides key metrics such as T-count, depths, and so on.

\subsubsection{Shell interoperability}
Developers can integrate Qsyn with external quantum synthesis tools, enhancing workflow flexibility through employing scripts, as detailed in Section~\ref{subsection:functionalities:usage}. This shell interoperability becomes particularly useful when comparing optimization results between Qsyn and other synthesis tools and when verifying optimization results with a third-party verifier.
For instance, a user could link multiple quantum computing tools in a chain with Qsyn, transforming input circuits and sending them to other tools for further operations, such as simulation or verification.

\subsection{Implementation advantage}
At the core of Qsyn CLI lie numerous data representations and utilities designed for convenient and efficient manipulations. The following discussion explores the critical design choices that lay the foundation for the QCS algorithm implementations in Qsyn.

\subsubsection{An extensible quantum circuit model}
Qsyn's quantum circuit data structure is highly expressive. Besides providing fundamental gate types, we allow developers to add new types of quantum gates freely. This expressivity, inspired by Qiskit's quantum circuit model, enables developers and researchers to combine a vast range of quantum operations, simplifying the management of these complex operations.

\subsubsection{More performant ZX-based optimization}
Qsyn's ZX-calculus-based synthesis routine results in quantum circuits with fewer two-qubit gates than PyZX's implementation by revising their algorithms in two aspects. The first is to adopt a more compact ZX-diagram representation for MCT gates and an early-stopping strategy to avoid producing redundant graph complexity~\cite{lu_dynamic_2023}. The other is to improve the conversion routine from ZX-diagrams to circuits. We have improved the ``gadget removal'' strategy during conversion to extract fewer $CZ$ gates, contributing to a more compacted quantum circuit. We detail the experimental data in Section~\ref{section:experiments}.

\subsubsection{Unified representation for Pauli rotation- and phase polynomial-based optimizations}
Qsyn's implementation of Pauli rotation- and phase polynomial-based QCS algorithms offers a significant advantage by using a unified Tableau-based representation for both approaches. This unification eliminates frequent data conversions required when switching between these two synthesis paradigms. Thus, Qsyn can provide hybrid synthesis strategies that leverage the strengths of both Pauli rotation-based and phase polynomial-based techniques, potentially leading to more optimized quantum circuits.

\subsection{Expediting community engagement}
As an open-source tool, Qsyn is committed to empowering a robust community for developers and researchers. We handle communications, issues, and feature requests on GitHub to foster a dynamic conversation environment for Qsyn's ongoing development. Furthermore, we adopt the following strategies to create a stable and inclusive development ecosystem:

\subsubsection{Cross-platform stability}
We aim to guarantee consistent results across different platforms, ensuring stability and uniform behavior regardless of the development environments. For example, the iteration order in the C++ standard hash maps (\texttt{std::unordered\_map}) varies across implementations, which is problematic when stable traversal orders are required. To address this, we have implemented an \texttt{ordered\_hashmap} in Qsyn to ensure stable traversals while maintaining fast access speeds.

\subsubsection{Docker distribution}
We provide a Docker environment to simplify Qsyn distribution and enhance the flexibility of development environments. This helps developers quickly replicate systems and troubleshoot environment-specific errors, leading to a more consistent user experience. It also supports development settings that might otherwise be incompatible.

\subsubsection{Code quality assurance}
Qsyn's CI/CD pipeline is pivotal in ensuring a reliable platform for development and research. Our project integrates an automated testing flow to verify software effectiveness and enforce code quality measures such as linting and formatting. This maintains high standards of code clarity and manageability, allowing contributors to focus more effectively on innovation. These measures are designed to enhance community involvement and contributions, cultivating an environment of collaboration and innovation vital to continuous evolution in Qsyn.

\section{Case study: Implementing a T-count optimization algorithm} \label{section:casestudy}
This section showcases how developers can design and implement a QCS algorithm with the Qsyn framework. Our exemplar algorithm is TODD, a T-count optimization algorithm for Clifford+$T$ circuits. We will demonstrate how to accelerate developing, verifying, and benchmarking the algorithm. 


TODD is an algorithm that can further optimize the $T$-count of the logical quantum circuit~\cite{heyfron_efficient_2018}. It converts the input circuit to an initial Clifford operator and a phase polynomial with no Clifford terms. Then, using an extended Lempel's matrix factorization algorithm, TODD obtains phase polynomials with fewer terms and corrects the Clifford difference before converting the polynomial back to a quantum circuit.

In the following, we will leverage Qsyn's Tableau data structure to build a new command for TODD. A Tableau represents a sequence of Clifford operators or Pauli rotation groups. Many algorithms can be implemented in terms of Tableaux, including the phase-merging~\cite{zhang_optimizing_2019} and internal-$H$-gate optimizations~\cite{vandaele_optimal_2024}. For our TODD algorithm, we will represent phase polynomials as a list of diagonal Pauli rotations, which is isomorphic to a phase polynomial. This type-punning avoids costly data representation conversions when designing synthesis routines that concatenate multiple algorithms. 

\subsection{Designing TODD usage flow with existing and new Qsyn commands}
Qsyn already supports many subroutines of TODD. The initial and final conversion between quantum circuits and the Tableau representation is available. Also, \texttt{tabl opt tmerge} can remove duplicated rotations, and \texttt{tabl opt hopt} can restrict the elements in Tableaux to only Clifford operators and phase polynomials. The remaining task is implementing the core algorithm, extended Lempel's algorithm. 

We position TODD as one of the phase polynomial optimization strategies and register it as a subcommand of the \texttt{tabl opt} command, which consists of all optimization algorithms for Tableaux. These considerations prompt us to design the following command interface:

\begin{center}
    \texttt{tabl opt phasepoly [strategy=todd]} 
\end{center}
Such an interface provides room for future additions of other optimization strategies. In Qsyn, a command consists of its name, the definition of its argument parser, and the action after the parse succeeds. We figuratively demonstrate the command implementation as follows, simplifying part of the code for readability:
\begin{lstlisting}[language=c++]
Command phasepoly_opt_cmd(TableauMgr& tabl_mgr) {
  return {"phasepoly",
          [](ArgumentParser& parser) { // parser definition
            parser.add_argument<std::string>("strategy")
              .default_value("todd")
              .choices({"todd"/*, more options... */}));
            },
          [&](ArgumentParser const& parser) { // on-success
            auto strategy =  // retrieve parse results
              parser.get<std::string>("strategy");
            if (strategy == "todd")
              optimize_phase_polynomial(
                *tabl_mgr.get(), 
                ToddStrategy{});
            return CmdExecResult::done;
          }};
}
\end{lstlisting}
For this command, the parser has only one optional argument for the \texttt{strategy} name, with \texttt{default\_value(...)} setting the default strategy TODD, and \texttt{choices({...})} constraining valid options (Line 6-8). Developers can use these attributes to ensure that the user provides valid arguments for the corresponding parameter, reducing tedious manual checks.

When parsing succeeds, we optimize the phase polynomials in the Tableau with the specified strategy. Using the parsing result retrieved with \texttt{parser.get<T>(...)} (Lines 12-13), we invoke the corresponding optimization algorithm. Finally, we add the command as a subcommand to \texttt{tabl opt}:
\begin{lstlisting}[language=c++]
Command tableau_opt_cmd(TableauMgr& tabl_mgr) {
  auto cmd = ...; // parent command definition
  ...;            // add other subcommands
  cmd.add_subcommand(phasepoly_opt_cmd(tabl_mgr));
  return cmd;
}
\end{lstlisting}

\subsection{Implementing the core algorithm}
Qsyn aims to empower QCS researchers by providing powerful and convenient constructs for algorithm development. Below, we illustrate this by detailing the core implementation of the TODD algorithm. The top-level call to the TODD core is figuratively shown as follows:

\begin{lstlisting}[language=c++]
auto todd(Clifford cliff, Polynomial poly) {
  properize(cliff, poly);
  auto const backup_poly = poly;
  auto const n_terms = poly.size();
  do {
    n_terms = poly.size();
    poly = todd_once(poly);
  } while (!poly.empty() && poly.size() < n_terms);
  apply_clifford_correction(cliff, backup_poly, poly);
  return {cliff, poly};
}
\end{lstlisting}
We have again simplified the code snippet for clarity. The implementation of this core routine demonstrates several facets of Qsyn's capabilities. First, the \texttt{properize} subroutine absorbs Clifford terms in the polynomial (\texttt{poly}) into the Clifford operator (\texttt{cliff}). It can be easily implemented by repurposing the available phase-merging optimization algorithm in~\cite{zhang_optimizing_2019}, an excellent example of how existing algorithms can be the cornerstone for further development.

The implementation continues with the \texttt{todd\_once} subroutine, which executes the crucial computations to minimize the $T$-count. This process leverages Qsyn's \texttt{BooleanMatrix} class, which supports calculations over Galois Fields, a common theme in QCS. Besides TODD, the \texttt{BooleanMatrix} is also used in the synthesis of Clifford gates, conversion from ZX-diagrams to quantum circuits, and many more. Finally, we calculate the Clifford correction by comparing the original and optimized polynomials. This is done by calculating a signature function detailed in \cite{heyfron_efficient_2018}.

We have seen how Qsyn provides convenient data structures and utility algorithms that developers can conveniently leverage. Ultimately, these utilities in Qsyn accelerate the development cycle and foster an environment conducive to breakthroughs in quantum computing research.

\subsection{Verifying the functionality}
We utilized numerical evaluations to verify that the algorithm can correctly optimize the phase polynomials for circuits with a small number of qubits:
\begin{lstlisting}[language=c++]
//!ARGS INPUT
qc read ${INPUT}
// TODD routine
convert qc tabl
tabl opt tmerge    // phase-merging optimization
tableau opt hopt   // to Clifford and phasepoly
tabl opt phasepoly // extended Lempel's Alg
convert tabl qc
// numerical verification
qc equiv --tensor
\end{lstlisting}

The listed commands above correspond to each step of the TODD routine and the equivalence checking of the two circuits by comparing their tensor form. For larger circuits, we tried reducing the original circuit concatenated with its adjoint to identity by applying our TODD implementation:
\begin{lstlisting}[language=c++]
// Read circuit and run TODD routine...
qc adjoint
qc compose 0 // compose the original circuit
// TODD routine...
tabl print   // check if the Tableau becomes empty
\end{lstlisting}


\subsection{Benchmarking the TODD algorithm}
We compared our TODD implementation with the other QCS approaches implemented in Qsyn. We utilized two particularly helpful commands: the \texttt{usage} command to display the memory and time usage and the \texttt{qc print --stat} command to collect the circuit statistics.

To showcase further how this new algorithm broadens Qsyn's optimization capabilities, we integrated the TODD algorithm into the Tableau-based synthesis routine. This routine, named \texttt{qtablq} as described in Section~\ref{subsection:functionalities:functionalities}, repeatedly applies phase-merging, internal-$H$-gate optimization, and TODD until the $T$-count converges. 

We separately performed this Tableau-based gate-level synthesis and its ZX-calculus-based counterparts on the exemplar \emph{hwb10} and \emph{urf2} circuits. Additionally, we ran a composite routine where the output circuit of the Tableau-based routine is fed to the ZX-calculus-based routine.

It is crucial to note that employing multiple synthesis pathways can further enhance the optimization objectives. As detailed in Table~\ref{tab:compare-diff-routes}, the pure Tableau method yielded circuits with fewer $T$-gates but with a notable presence of two-qubit gates, resulting in deeper circuits. Conversely, the ZX-calculus method was not as powerful in eliminating $T$-gates but effectively compressed circuit depth. However, combining these methods enabled the simultaneous optimization of both circuit depth (or two-qubit gates) and the number of rotations. 

\begin{table}[!t]
    \centering
    \vspace{-10px}
    \caption{Comparison of different synthesis routes in Qsyn}
    \begin{tabular}{ll|rrrrr}
    \toprule
        Circuit&Method&Clifford & \#H & 2-qubit & \#T & Depth \\ \midrule\midrule
        \multirow{3}{*}{hwb10} & Tableau & 218159 & 5208 & 208416 & 15681 & 414246 \\ 
        ~ & ZX & 53222 & 5248 & 45381 & 15891 & 82107 \\ 
        ~ & Tabl+ZX & 52515 & 5244 & 44986 & 15681 & 79893 \\ \midrule
         \multirow{3}{*}{urf2} & Tableau & 39039 & 1704 & 36079 & 5007 & 74274 \\ 
        ~ & ZX & 11375 & 1716 & 9307 & 5063 & 18598 \\ 
        ~ & Tabl+ZX & 11955 & 1704 & 9766 & 5007 & 19267 \\ \bottomrule
    \end{tabular}
    \label{tab:compare-diff-routes}
    \vspace{-10px}
\end{table}

\section{Experiments} \label{section:experiments}
We evaluate Qsyn's performance with a three-part experiment: first, we compare Qsyn's efficiency and optimization power with other existing synthesis frameworks, including Qiskit~\cite{aleksandrowicz_qiskit_2019}, \tket~\cite{sivarajah_tket_2021}, and PyZX~\cite{kissinger_pyzx_2020}; second, we compare the Qsyn's mapping with additional mapping framework including QMAP~\cite{wille_mqt_2023}, SABRE~\cite{li_tackling_2019} and 
TOQM~\cite{zhang_time-optimal_2021}; then, we demonstrate how the multiple synthesis strategies Qsyn provides work in addressing the various optimization objectives.

\subsection{Settings}
We conducted our experiments on an Ubuntu 22.04 workstation with a 5.5GHz Intel\textsuperscript{\textregistered} Core\textsuperscript{\texttrademark} i9-13900K CPU and 128GB of RAM. We set time limits of one day and memory limits of 128GB for our experiments. If these limits were exceeded, we denoted the outcomes as TLE/ MLE (Time/Memory Limit Exceeded). Regarding circuit statistics, we computed the $R_Z$-count (referred to as $\#R_Z$ in the subsequent tables), as well as the depth of the circuits. We calculated the gate delays with the methodology outlined in~\cite{deng_codar_2020}, counting single-qubit operations as 1 unit and two-qubit operations as 2 units.

We opted for using \emph{hwb} (hidden weighted bit), \emph{urf} (unstructured reversible function), \emph{ham} (Hamming code), \emph{gf} (Galois field), \emph{adder},  \emph{qugan}, and \emph{multiplier} as benchmark circuits. These benchmarks encompass circuits commonly encountered in gate-level synthesis and mapping tasks. The synthesis set is selected from PyZX~\cite{kissinger_pyzx_2020}, while the mapping set is a subset of circuits from RevLib~\cite{wille_revlib_2008} collected by QMAP~\cite{wille_mqt_2023}. Additionally, we included a selection of circuits tailored for NISQ devices, as recently proposed in QASMBench \cite{li_qasmbench_2023}.

It is worth noting that as Qsyn is a growing framework, its performance is expected to be improved continuously. The following data represents the collective efforts of the framework's contributors, some of them being the authors of this paper, at the time of writing.

\subsection{Comparison with other frameworks on gate-level synthesis}

We performed gate-level synthesis to examine the run time and memory usage of Qsyn against other frameworks in Table \ref{tab:compare-other-frameworks-syn}. Table~\ref{tab:methods} lists the methods we employed and the corresponding function calls. The output gate set was Clifford+$R_Z$\footnote{Here, the $R_Z$ mostly refers to $T$ gates. However, cases like \emph{ising} consist of smaller angles of $R_Z$ gates, so we collectively marked them as $R_Z$ gates.}. 
\begin{table*}[t]
    \scriptsize
    \centering
    \caption{Comparison with Other frameworks on the gate-level synthesis stage. Time unit: \textup{s}; Memory unit: \textup{MiB}}
    \label{tab:compare-other-frameworks-syn}
    \setlength\tabcolsep{4pt}
    \newcolumntype{Y}{>{\raggedleft\arraybackslash}X}
    \begin{tabularx}{\textwidth}{c|YYYY|YYYY|YYYY|YYYY}
\toprule
    \multicolumn{1}{c}{Circuit}&\multicolumn{4}{c|}{Qiskit \cite{aleksandrowicz_qiskit_2019}}&\multicolumn{4}{c|}{\tket \cite{sivarajah_tket_2021}}&\multicolumn{4}{c|}{PyZX \cite{kissinger_pyzx_2020}}&\multicolumn{4}{c}{Qsyn}\\
    \cmidrule{2-5}\cmidrule{6-9}\cmidrule{10-13}
    \cmidrule{14-17}
   \multicolumn{1}{c}{name} & 
   \multicolumn{1}{c}{$\#R_Z$}& \multicolumn{1}{c}{Depth} & \multicolumn{1}{c}{Time} & \multicolumn{1}{c|}{Mem} &
   \multicolumn{1}{c}{$\#R_Z$}& \multicolumn{1}{c}{Depth} & \multicolumn{1}{c}{Time} & \multicolumn{1}{c|}{Mem} &
   \multicolumn{1}{c}{$\#R_Z$}& \multicolumn{1}{c}{Depth} & \multicolumn{1}{c}{Time} & \multicolumn{1}{c|}{Mem} &
   \multicolumn{1}{c}{$\#R_Z$}& \multicolumn{1}{c}{Depth} & \multicolumn{1}{c}{Time} & \multicolumn{1}{c}{Mem} \\
        \midrule
        \midrule
         cm42 & 770 & 1578 & 1.55 & 372.4 & 770 & 1577 & 0.35 & 188.2 & \bf{370} & 1401 & 3.696 & 147.0 & \bf{370} & \bf{1385} & 0.10 & 4.25 \\ 
        cm85 & 4984 & 10632 & 7.16 & 372.4 & 4984 & 10630 & 5.56 & 218.8 & \bf{1950} & 10543 & 83.22 & 170.0 & \bf{1950} & \bf{9621} & 3.93 & 169.5 \\ 
        \midrule
        ham7 & 133 & 319 & 0.14 & 377.2 & 133 & 317 & 0.05 & 182.0 & \bf{81} & 301 & 0.31 & 139.3 & \bf{81} & \bf{268} & 0.01 & 1.25 \\ 
        ham15 & 2173 & 4700 & 4.49 & 377.2 & 2173 & 4993 & 0.91 & 194.8 & 1019 & \bf{4270} & 35.22 & 160.4 & \bf{1008} & 4281 & 10.71 & 182.0 \\ 
        \midrule
        hwb8 & 5437 & \bf{13238} & 15.05 & 392.5 & 5479 & 14180 & 3.89 & 219.5 & 3517 & 15197 & 113.6 & 210.7 & \bf{3460} & 15605 & 11.25 & 229.2 \\ 
        hwb9 & 90858 & 194125 & 179.2 & 383.1 & 90858 & 193852 & 5481 & 718.4 & 43572 & \bf{176425} & 5982 & 1142 & \bf{43565} & 185294 & 516.7 & 6090 \\ 
        hwb10 & 26915 & \bf{61520} & 77.52 & 349.8 & 26999 & 65491 & 199.4 & 563.3 & 15891 & 79892 & 2343 & 837.7 & \bf{15681} & 79893 & 1808 & 4880 \\ 
        hwb12 & 152755 & \bf{357195} & 432.0 & 562.3 & 153551 & 383206 & 5402 & 1112 & \multicolumn{4}{c|}{MLE} & \bf{85611} & 597419 & 4350 & 7471 \\ 
        \midrule
        rd73 & 2317 & 4835 & 3.38 & 372.4 & 2317 & 4818 & 1.63 & 203.7 & \bf{1045} & \bf{4463} & 15.27 & 160.9 & \bf{1045} & 4528 & 1.54 & 79.90 \\ 
        rd84 & 5957 & 12186 & 8.62 & 371.4 & 5957 & 12168 & 7.78 & 227.8 & \bf{2625} & 11953 & 69.43 & 196.4 & \bf{2625} & \bf{11583} & 8.25 & 572.1 \\ 
        \midrule
        sym9 & 15232 & 32071 & 30.03 & 357.4 & 15232 & 32061 & 62.83 & 322.8 & \bf{6450} & \bf{31485} & 373.0 & 270.2 & \bf{6450} & 33078 & 25.45 & 1299 \\ 
        sym10 & 28084 & 59290 & 55.32 & 358.3 & 28084 & 59286 & 205.5 & 357.8 & \bf{11788} & 60933 & 1247 & 429.7 & \bf{11788} & \bf{60485} & 114.4 & 892.1 \\ 
        \midrule
        urf2 & 7707 & 19705 & 18.25 & 374.4 & 7707 & 20006 & 20.22 & 245.2 & 5065 & \bf{18824} & 350.7 & 308.4 & \bf{5007} & 19267 & 6.45 & 175.6 \\ 
        urf4 & 224028 & 449853 & 202.9 & 612.6 & \multicolumn{4}{c|}{MLE} & \multicolumn{4}{c|}{MLE} & \bf{97036} & \bf{389621} & 9811 & 8709 \\ 
        urf6 & 75180 & 161026 & 67.54 & 412.9 & 75180 & 160813 & 3517 & 637.0 & 33026 & \bf{129584} & 10527 & 671.0 & \bf{31568} & 133269 & 6751 & 15902 \\ 
        \midrule
        adder\_433 & 2304 & \bf{3257} & 6.61 & 353.6 & 2304 & 3504 & 1.59 & 201.6 & \bf{1536} & 3802 & 43607 & 241.8 & \bf{1536} & 4298 & 19.5 & 213.2 \\ 
        adder\_64 & 336 & \bf{551} & 0.86 & 354.6 & 336 & 593 & 0.18 & 185.9 & \bf{224} & 670 & 41.57 & 243.8 & \bf{224} & 683 & 0.3 & 23.59 \\ 
        \midrule
        gf2\^{}128 & 98304 & \bf{6744} & 2929 & 298.4 & 98304 & 6870 & 68.31 & 638.5 & \multicolumn{4}{c|}{TLE} & \bf{65664} & 613170 & 1586 & 1802 \\ 
        gf2\^{}64 & 24576 & \bf{3356} & 371.8 & 180.8 & 24576 & 3418 & 11.25 & 312.9 & 16448 & 106078 & 23037 & 341.8 & \bf{16338} & 67854 & 284.3 & 408.2 \\ 
        gf2\^{}32 & 6144 & \bf{1660} & 49.73 & 150.2 & 6144 & 1690 & 2.29 & 215.4 & \bf{4128} & 18571 & 826.3 & 188.5 & \bf{4128} & 11718 & 9.34 & 68.77 \\ 
        gf2\^{}16 & 1536 & \bf{812} & 7.11 & 141.2 & 1536 & 826 & 0.55 & 189.3 & \bf{1040} & 5279 & 46.64 & 146.0 & \bf{1040} & 3267 & 1.08 & 12.74 \\ 
        \midrule
        ising\_420 & 1048 & 13 & 0.57 & 352.7 & 1048 & 13 & 0.07 & 182.2 & \bf{839} & 153381 & 15572 & 232.1 & \bf{839} & \bf{12} & 1.06 & 5.25 \\ 
        ising\_98 & 243 & 13 & 1.18 & 350.7 & 243 & 13 & 0.18 & 201.5 & \bf{195} & 9887 & 54.30 & 170.8 & \bf{195} & \bf{12} & 0.01 & 1.25 \\ 
        ising\_34 & 83 & 13 & 0.46 & 352.7 & 83 & 13 & 0.06 & 181.2 & \bf{67} & 997 & 0.93 & 138.7 & \bf{67} & \bf{12} & 0.01 & 1.00 \\ 
        ising\_16 & 262 & 75 & 0.65 & 350.7 & 262 & 74 & 0.15 & 201.5 & \bf{204} & 200 & 0.41 & 160.0 & \bf{204} & \bf{58} & 0.01 & 2.00 \\ 
        \midrule
        knn\_341 & \bf{1360} & 2223 & 44.48 & 152.6 & \bf{1360} & \bf{2222} & 9.41 & 243.6 & \bf{1360} & 3380 & 6.81 & 258.8 & \bf{1360} & 3404 & 19.86 & 37.21 \\ 
        knn\_129 & \bf{512} & 845 & 7.65 & 142.4 & \bf{512} & \bf{844} & 0.68 & 189.8 & \bf{512} & 1343 & 1255 & 261.7 & \bf{512} & 1282 & 0.99 & 3.00 \\ 
        \midrule
        mult\_350 & \bf{136080} & \bf{219465} & 2260 & 361.7 & \bf{136080} & 233645 & 355.0 & 1136 & \multicolumn{4}{c|}{TLE} & \multicolumn{4}{c}{TLE} \\ 
        mult\_45 & 2124 & 3475 & 9.72 & 445.4 & 2124 & 3687 & 0.88 & 191.4 & \bf{634} & \bf{2702} & 281.1 & 169.5 & \bf{634} & 2987 & 31.05 & 184.1 \\ 
        \midrule
        qugan\_395 & 2753 & 3759 & 3.41 & 358.6 & 3292 & \bf{2586} & 19.82 & 240.1 & \bf{2360} & 7092 & 46040 & 914.6 & 2361 & 3812 & 111.5 & 142.8 \\ 
        qugan\_111 & 765 & 1061 & 15.85 & 358.6 & 918 & \bf{740} & 0.91 & 193.2 & \bf{657} & 2326 & 568.9 & 215.9 & \bf{657} & 1114 & 1.65 & 8.75 \\ 
        \midrule
        \midrule
        Geo. Mean $\Delta$\footnotemark & 1.679 & 0.680 & 1.337 & 3.010 & 1.560 & 0.615 & 0.583 & 2.846 & 1.004 &	2.472 &	43.908 & 3.586 & 1.000 & 1.000 & 1.000 & 1.000\\
        \bottomrule
    \end{tabularx}
    \vspace{-10px}
\end{table*}

\begin{table}[t]
    \centering
    \caption{The methods and the corresponding function calls}
    \label{tab:methods}
    \footnotesize
    \begin{tabularx}{\linewidth}{lrX}
    \toprule
        Method & Ver. & Function(s) -- Default Pass(es)  \\
    \midrule\midrule
        Qiskit  & 1.0.1  & \texttt{generate\_preset\_pass\_manager}        \\
        PyZX    & 0.8.0  & \texttt{to\_graph, full\_reduce, extract\_circuit, basic\_optimization}             \\
        \tket{} & 1.27.0 & \texttt{KAKDecomposition, CliffordSimp, SynthesiseTket, auto\_rebase\_pass}        \\
    \bottomrule
    \end{tabularx}
    \vspace{-15px}
\end{table}

In Table~\ref{tab:compare-other-frameworks-syn}, the default Qiskit and \tket{} synthesis methods tended to produce circuits in shorter times and excelled at generating shallower circuits. However, the $R_Z$-counts were typically much larger than those of PyZX and Qsyn.

Compared to PyZX, which also prioritizes minimizing small-angle rotations, Qsyn tended to optimize circuits with shorter program runtimes and lower memory usage. This distinction was particularly notable in circuits such as \emph{gf2\^{}128}, \emph{hwb12}, and \emph{urf4}, where PyZX encounters TLE and/or MLE. For circuit statistics, Qsyn achieved either smaller or equal $R_Z$-counts and, on average, shallower circuits. This is attributed to Qsyn's integration of ZX-calculus and Tableau combined methods, further compressing rotation counts. 

Furthermore, for circuit depths, in cases with numerous non-Clifford rotation gates, like \emph{ising} and \emph{qugan}, the ZX-calculus-based method tended to over-perform its transformation rules, resulting in excessive edges and deeper circuits. In sum, Qsyn generated better circuits in shorter times and with lower resource consumption when minimizing $R_Z$-count. The experimental results demonstrate Qsyn's competitiveness over the state-of-the-art frameworks, underscoring the quality of its supporting methods and code infrastructure.\footnotetext{When calculating the normalized ratios, we omitted the cases if baseline Qsyn or the framework itself got a TLE or MLE.}

\begin{table*}[t]
\caption{Performance on various large benchmarks. The boldface results represent the best ones among QMAP, Qiskit, \tket{}, and SABRE. Cost is the depth of the mapped circuit with $\Delta Cost = \frac{Cost_{Qsyn}}{Cost_{best}}$ and $\Delta Time = \frac{Time_{Qsyn}}{Time_{best}}$. Time unit: \textup{s}}
    \centering
    \scriptsize
    \setlength\tabcolsep{4pt}
    \newcolumntype{Y}{>{\raggedleft\arraybackslash}X}
    \begin{tabularx}{\textwidth}{cYYYYYYYYYY|YYYY}
\toprule
    \multicolumn{3}{c}{Original Circuit}&\multicolumn{2}{c}{Qiskit \cite{aleksandrowicz_qiskit_2019}}&\multicolumn{2}{c}{QMAP \cite{wille_mqt_2023}}&\multicolumn{2}{c}{\tket \cite{sivarajah_tket_2021}}&\multicolumn{2}{c}{SABRE \cite{li_tackling_2019}} &\multicolumn{4}{|c}{Qsyn}\\
\cmidrule{4-5}\cmidrule{6-7}\cmidrule{8-9}\cmidrule{10-11}\cmidrule{12-15}
   \multicolumn{1}{c}{name} & \multicolumn{1}{c}{$\#$Qubit}& \multicolumn{1}{c}{Depth} & \multicolumn{1}{c}{Cost} & \multicolumn{1}{c}{Time}  & \multicolumn{1}{c}{Cost} & \multicolumn{1}{c}{Time}& \multicolumn{1}{c}{Cost} & \multicolumn{1}{c}{Time} &\multicolumn{1}{c}{Cost} & \multicolumn{1}{c}{Time} & \multicolumn{1}{|c}{Cost} & \multicolumn{1}{c}{Time} &\multicolumn{1}{c}{$\Delta$ Cost} &\multicolumn{1}{c}{$\Delta$ Time}\\
  \midrule
         \multirow{4}{*}{gf} & 384 & 7762 & \bf{87,893} & 654.1 & 375,442 & 21.57 & 393,352 & 906.8 & 102,912 & \bf{5.407} & 65,523 & 3.020 & 0.745 & 0.559 \\
        ~ & 192 & 3862 & \bf{36,278} & 95.10 & 69,094 & 5.332 & 138,102 & 79.53 & 41,825 & \bf{1.352} & 27,493 & 1.240 & 0.758 & 0.917 \\ 
        ~ & 96 & 1910 & 16,156 & 6.791 & 23,243 & 1.086 & 28,644 & 7.239 & \bf{15,643} & \bf{0.333} & 11,931 & 0.690 & 0.763 & 2.072 \\ 
        ~ & 48 & 934 & 5,029 & 0.818 & \bf{4,867} & 0.238 & 9,299 & 1.331 & 4,943 & \bf{0.110} & 3,988 & 0.420 & 0.819 & 3.818 \\ 
        \midrule
         \multirow{2}{*}{adder} & 433 & 3839 & 16,228 & 42.95 & \bf{14,559} & 6.788 & 30,906 & 150.3 & 17,693 & \bf{0.255} & 9,083 & 0.790 & 0.624 & 3.098 \\ 
        ~ & 64 & 641 & 2,492 & 0.851 & 1,747 & 0.066 & 2,482 & 1.725 & \bf{1,684} & \bf{0.024} & 1,636 & 0.170 & 0.917 & 7.083 \\ 
        \midrule
         \multirow{2}{*}{knn} & 341 & 2225 & 11,258 & 7.823 & 9,664 & 0.729 & 22,894 & 15.04 & \bf{9,127} & \bf{0.114} & 6,823 & 0.880 & 0.748 & 7.719 \\ 
        ~ & 129 & 847 & 3,750 & 2.965 & 3,518 & 1.058 &  8,317 & 2.754 & \bf{3,170} & \bf{0.103} & 2,458 & 0.890 & 1.290 & 8.641 \\ 
        \midrule
         \multirow{2}{*}{qugan} & 395 & 3370 & 18,879 & 26.80 & \bf{11,362} & 2.274 & 22,442 & 469.7 & 26,104 & \bf{1.328} & 8,059 & 0.980 & 0.709 & 0.738 \\ 
        ~ & 111 & 616 & 4,850 & 1.896 & \bf{3,193} & 0.204 & 3,792 & 5.788 & 4,270 & \bf{0.063} & 2,453 & 0.270 & 1.302 & 4.286 \\ 
        \midrule
         \multirow{2}{*}{multiplier} & 350 & 233656 & 963,091 & 964.5 & TLE & TLE & 1,904,264 & 6060 & \bf{837,539} & \bf{5.673} & 627,440 & 5.260 & 0.749 & 0.927 \\ 
        ~ & 45 & 3692 & 13,787 & 1.720 & \bf{10,845} & 0.365 & 13,121 & 1.885 & 11,769 & \bf{0.103} & 10,256 & 0.790 & 0.946 & 7.670 \\ 
        \midrule
        \midrule
        \multicolumn{3}{c}{Geometric Mean $\Delta$}&& ~ & ~ & ~ & ~ & ~ &  &&&& 0.776 & 2.674 \\
      \bottomrule
      \end{tabularx}
    \vspace{-10px}
    \label{table:large-scale}
\end{table*}

\subsection{Comparison with other frameworks on qubit mapping}

To assess the runtime and effectiveness of Qsyn compared to other frameworks, we conducted qubit mapping on \emph{urf} and \emph{hwb} benchmarks with less than 20 qubits, as well as QASMBenchmark circuits ranging from 45 to 433 qubits. To ensure a fair comparison, we utilized the same transpiled circuits as input, consisting solely of Clifford and $R_Z$ gates. 

We employed IBM Qiskit 1.0.1 with both the default and SABRE mappers and the default heuristic versions of QMAP, \tket, and TOQM. In both Qsyn and the original Duostra framework~\cite{cheng_robust_2024}, we used the search-scheduler for the \emph{urf} and \emph{hwb} tasks, while employing the heuristic-scheduler for larger cases. As for the target topologies, we opted for the IBMQ heavy-hexagon device with the minimum number of physical qubits that could accommodate the logical circuit's qubits. 

\begin{figure}
    \centering
    \includegraphics[width=\linewidth]{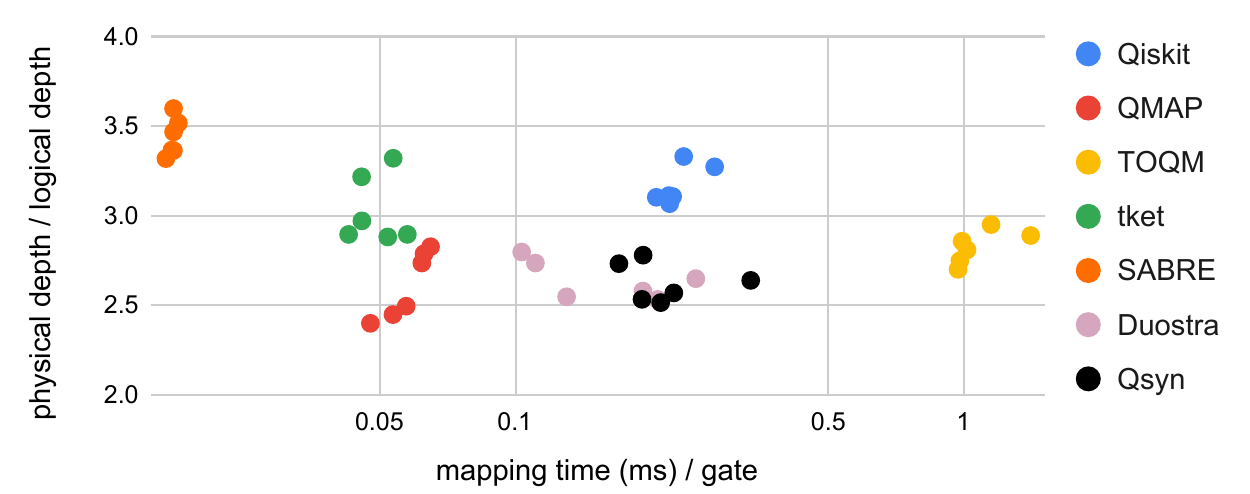}
    \vspace{-22px}
    \caption{Results of different mapping frameworks on \emph{urf} circuits}
    \label{fig:urf-mapping}
\end{figure}

\begin{figure}
    \centering
    \includegraphics[width=\linewidth]{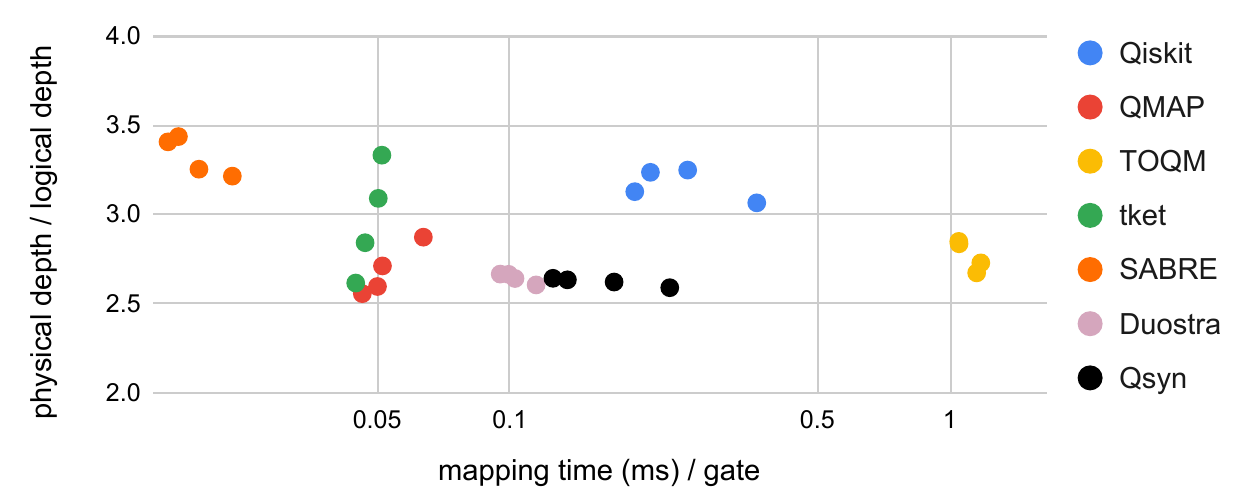}
    \vspace{-22px}
    \caption{Results of different mapping frameworks on \emph{hwb} circuits}
    \vspace{-10px}
    \label{fig:hwb-mapping}
\end{figure}

Fig.~\ref{fig:urf-mapping} and Fig.~\ref{fig:hwb-mapping} depict the mapping results on the 16-qubit device. SABRE exhibited the shortest mapping time per gate of all mappers, followed by \tket{} and QMAP. The original Duostra, Qsyn, and Qiskit were slightly slower, routing with a 0.5 ms/gate rate. In contrast, it took the TOQM mapper almost 1 ms to route a gate. Regarding the routing quality, the QMAP, Qsyn, Duostra, and TOQM mappers typically result in a 2.7x increase, while \tket{} showed a wide range of mapping overheads. On the other hand, SABRE and Qiskit exhibited a more-than-3x increase in depths. Furthermore, the mapping results of Qsyn and Duostra were nearly identical. Qsyn's version took slightly longer runtime because Qsyn's framework additionally supports extensible gate types, resulting in data access overheads, while the original Duostra implementation only supports a small set of primitive quantum gates. These results demonstrate that the research results of~\cite{cheng_robust_2024} can be elegantly replicated within the Qsyn framework. 

We list the mapping outcomes for larger circuits in Table~\ref{table:large-scale},  emphasizing the best result (marked as bold) among the four methods under comparison for each scenario. Subsequently, we calculate our enhancements relative to these optimal outcomes. The data indicates that, on average, Qsyn achieved 22.4\% improvement over the best performance among the compared methods, showcasing that Qsyn supports a performant mapper for quantum circuits of various scales.

\section{Conclusion} \label{section:conclusion}
In this paper, we discussed the development of Qsyn and described its basic usage. We discussed how it can speed up the workflow of developers and researchers in the QCS field and demonstrated the framework's capability and competiveness with a detailed case study and thorough experiments.

Qsyn is a thriving platform with new features being added regularly. Some of the future work for our framework includes:
\subsubsection{Improving high-level synthesis}
Even with powerful gate-level optimization algorithms, the quality of high-level quantum circuits remains crucial to the synthesis result. A well-designed high-level synthesis algorithm can yield efficient high-level circuits and adapt to the qubit resource constraint, simplifying the subsequent synthesis steps.
\subsubsection{More flexible gate-level synthesis}
We have seen in Section~\ref{section:casestudy} and \ref{section:experiments} how multiple synthesis paradigms can help generate better circuits. We aim to optimize multiple metrics simultaneously and characterize the trade-off between them. Device-aware synthesis strategies are also promising for further compacting the resulting circuits.

\subsubsection{Beyond NISQ devices} 
Fault-tolerant quantum architectures require decomposing the Clifford+$T$ gate set and introducing new Clifford gate models. Distributed devices are also a common theme for realizing quantum advantage, necessitating algorithms targeting specifically for these devices.


\bibliographystyle{ieeetr}
\bibliography{Qsyn-QCE}

\end{document}